\documentclass[10pt]{article}
\usepackage{amssymb,amsmath,euscript,epsfig}
 
\newcommand{\be}[1]{\begin{equation}\label{#1}}
\newcommand{\ee}{\end{equation}}
\newcommand{\ba}[1]{\begin{eqnarray}\label{#1}}
\newcommand{\ea}{\end{eqnarray}}
\newcommand{\rf}[1]{(\ref{#1})}
\newcommand{\nn}{\nonumber}
\newcommand{\etal}{{\it et al }}

\def\RR{{\mathbb R}}

\def\CC{{\mathbb C}}

\def\ZZ{{\mathbb Z}}

\def\Ai{\mbox{Ai}}

\newcommand{\fK}{\cal{K}}
\newcommand{\vect}[1]{\mathbf{#1}}
\newcommand{\bsym}[1]{\boldsymbol{#1}}
\newcommand{\const}{\mbox{\rm const}}

\begin{document}
\title{MHD $\alpha^2-$dynamo, Squire equation and ${\cal PT}-$symmetric interpolation between
square well and harmonic oscillator}
\author{Uwe G\"unther$^a$\footnote{e-mail:
u.guenther@fz-rossendorf.de}\, , Frank
Stefani$^{a}$\footnote{e-mail: f.stefani@fz-rossendorf.de}\, and
Miloslav Znojil$^b$\footnote{ e-mail: znojil@ujf.cas.cz}
\\[2ex]
$^a$ Research Center Rossendorf, P.O. Box 510119, D-01314 Dresden, Germany\\[1ex]
$^b$ \'{U}stav jadern\'e fyziky AV \v{C}R, 250 68 \v{R}e\v{z},
Czech Republic}
\date{27 January 2005}

\maketitle
\begin{abstract}
It is shown that the $\alpha^2-$dynamo of Magnetohydrodynamics, the
hydrodynamic Squire equation as well as an interpolation model of
${\cal PT}-$symmetric Quantum Mechanics are closely related as
spectral problems in Krein spaces. For the $\alpha^2-$dynamo and the
${\cal PT}-$symmetric model the strong similarities are demonstrated
with the help of a $2\times 2$ operator matrix representation,
whereas the Squire equation is re-interpreted as a rescaled and
Wick-rotated ${\cal PT}-$symmetric problem. Based on recent results
on the Squire equation the spectrum of the ${\cal PT}-$symmetric
interpolation model is analyzed in detail and the Herbst limit is
described as spectral singularity.
\end{abstract}



\section{Introduction}
Non-Hermitian ${\cal PT}-$symmetric quantum mechanical systems
\cite{BB,BBjmp,PT-Z1,most1,most3,BBJ-1,BBJ-a,B-cz1} are known to
possess spectral sectors with purely real eigenvalues as well as
sectors with pairs of complex conjugate eigenvalues. Changes of
certain system parameters can lead to spectral phase transitions
from one sector to the other. The physics in the two sectors has
been identified with phases of unbroken ${\cal PT}-$symmetry (real
eigenvalues) and spontaneously broken ${\cal PT}-$symmetry (pairwise
complex conjugate eigenvalues) \cite{BB,BBjmp}. From a mathematical
point of view, non-Hermitian ${\cal PT}-$symmetric Hamiltonians are
self-adjoint operators in Krein spaces \cite{azizov,L2,LT-1}
--- Hilbert spaces with an additional indefinite metric structure
--- and the two spectral sectors correspond to Krein space states
of positive or negative type (real eigenvalues) and neutral
(isotropic) states (pairwise complex conjugate eigenvalues).

Apart from ${\cal PT}-$symmetric Quantum Mechanics (PTSQM), it is
known that a certain class of spherically symmetric mean-field
dynamo models \cite {krause1} of Magnetohydrodynamics (MHD) can be
described by self-adjoint operators in Krein spaces as well
\cite{GS-jmp1}. These models show similar spectral phase transitions
from real to pairwise complex conjugate eigenvalues \cite{GSG-cz2}
--- and only the physical interpretation differs from that in PTSQM.
For dynamos it simply consists in a transition from non-oscillatory
states to oscillatory states.

In the present paper, we are going to briefly describe the
underlying structural operator theoretic parallels between PTSQM
models and the spherically symmetric MHD $\alpha^2-$dynamo (section
\ref{PTSQM}). The discussion will be illustrated with the help of a
${\cal PT}-$symmetric interpolation between a harmonic oscillator
placed in a square well and an empty square well
 (section \ref{interpol}). This interpolation shows a rich
structure of spectral phase transitions with a couple of unexpected
features. Furthermore, we will show in sect. \ref{Herbst-box} that
the eigenvalue problem of the ${\cal PT}-$symmetric (intermediate)
interpolation model with linear complex potential (purely complex
electrical field) within the square well is mathematically identical
to the eigenvalue problem of the rescaled and Wick-rotated Squire
equation of hydrodynamics which describes the normal vorticity of a
plane channel flow (Couette flow) with linear transversal velocity
profile. Recent Airy function based results on the Squire equation
allow us to analytically describe the spectral behavior of the PTSQM
model in the limiting case when the width of the square well tends
to infinity. In this limit we reproduce the Herbst model
\cite{Herbst-1} with its empty spectrum. The limiting behavior
occurs as a blowing-up of the spectrum to infinity along three
directions on the complex plane --- leaving behind a spectrally
empty region at any fixed finite distance from the origin of the
spectral plane. In section \ref{conclu} we briefly sketch some links
of the obtained results to other physical setups and analytical
techniques.

\section{Krein space properties of ${\cal PT}-$symmetric quantum models
and of the spherically symmetric MHD $\alpha^2-$dynamo\label{PTSQM}}

\subsection{${\cal PT}-$symmetric quantum models}

In their seminal letter \cite{BB} Bender and Boettcher identified
${\cal PT}-$symmetry as the essential property of the
non-Hermitian quantum system
\be{BBeq}
 H\psi(x) = E\,\psi(x), \ \ \ \ \ \ \ \ \ \ \ \ \ \
 H = -\frac{d^2}{dx^2} + g\,x^2(ix)^\nu
\ee
which ensures the reality of its spectrum for exponents $\nu \in [0,
2)$ and $\psi(x) \in \tilde{\cal H}=L_2(-\infty,\infty)$
\cite{DDT-1}. This allowed them not only to extend an earlier
conjecture of Bessis and Zinn-Justin (whose numerical results
indicated that quantum systems with complex potential $V(x)=ix^3$
might have a purely real spectrum), but also initiated the still
lasting intensive study of generalized ${\cal PT}-$symmetric
non-Hermitian systems \cite{CMB-rev}. Such systems are characterized
by a ${\cal PT}-$symmetric Hamiltonian $H$,
\be{i1}
[{\cal PT},H]=0
\ee
where ${\cal P}$ denotes a reflection
\be{t2}
{\cal P}x{\cal P}=-x,\qquad {\cal P}\psi (x)=\psi(-x)
\ee
while the time-reversal operator ${\cal T}$ performs complex
conjugation
\be{t1}
{\cal T}i{\cal T}=-i,\qquad {\cal T}\psi (x)=\psi(x)^*.
\ee
Because both operators ${\cal P}$ and ${\cal T}$ are involution
operators,
\be{t3}
{\cal P}^2=I,\qquad {\cal T}^2=I,
\ee
they induce natural $\ZZ_2-$gradings of the Hilbert space $\tilde
{\cal H}$. For our subsequent analysis it suffices to consider the
subclass of models which can be defined solely over the real line
$x\in \RR$. For such models the ${\cal T}-$induced $\ZZ_2-$grading
corresponds to a splitting of the wave functions $\psi \in \tilde
{\cal H}$ into real and imaginary components (what is of no direct
physical interest in a quantum mechanical context; additionally one
would have to work in a real Hilbert space with doubled dimension
compared to the original complex one), whereas ${\cal P}$ induces a
$\ZZ_2-$grading into parity even and parity odd components
\be{t4} \psi(x)=\psi_+(x)
+\psi_-(x),\qquad {\cal P} \psi_\pm(x)=\psi_\pm(-x)=\pm
\psi_\pm(x)\, .
\ee
The corresponding $\ZZ_2-$graded Hilbert space splits as
\be{t4a}
\tilde {\cal H}={\cal H}_+\oplus {\cal H}_-, \qquad \psi_\pm \in
{\cal H}_\pm\, .
\ee

In the case of a simple ${\cal PT}-$symmetric one-particle system
with Hamiltonian
\be{t5}
H=-\partial^2_x+V_+(x)+iV_-(x), \qquad V_\pm (-x)=\pm V_\pm
(x),\quad \Im V_\pm =0
\ee
it holds
\be{t6}
H={\cal P}H^\dagger{\cal P}
\ee
and ${\cal P}$ is a so called fundamental (canonical) operator
symmetry \cite{azizov,L2} of $H$ --- i.e.,  $H$ is ${\cal
P}-$pseudo-Hermitian in the sense of Refs. \cite{most1,most3}.
Operators with an involutive fundamental symmetry are known to be
symmetric
--- and for appropriately chosen domain (boundary conditions for the
functions $\psi(x)$) even self-adjoint
--- in a Krein space $\fK$. For ${\cal P}-$pseudo-Hermitian
operators over the real line this Krein space $\fK_{\cal P}$ is
given as \cite{LT-1,MZ-sep,japar}
\be{t7}
\left(\fK_{\cal P},[.,.]_{\cal P}\right),\qquad [\psi,\phi]_{\cal
P}=(\psi,{\cal P}\phi)=\int_{C\subseteq \RR} \psi^*(x){\cal
P}\phi(x)d x=\int_{C\subseteq \RR} \psi^*(x)\phi(-x)d x\, .
\ee
Depending on the concrete problem, the integration in \rf{t7} is
performed over a finite interval, $C=[-a,a]$, or over the complete
real line, $C=(-\infty,\infty)\sim \RR $. {}From \rf{t5}, \rf{t6}
one immediately finds
\be{t8}
[H\psi,\phi]_{\cal P}=[\psi,H\phi]_{\cal P}\, .
\ee
 The Krein space inner (scalar) product $[\psi,\phi]_{\cal P}$ has
the following properties:
\begin{itemize}
\item It  coincides with the more general ${\cal
PT}$ inner product of C. Bender \etal \cite{BBJ-1,BBJ-a}
\be{t9}
(\psi,\phi)_{\cal PT}=\int_{C\subset \CC} [{\cal
PT}\psi(x)]\phi(x) dx
\ee
when the integration path $C\subset \CC$ of the latter integral is
restricted to (an interval of) the real line, $C\subseteq \RR $,
\be{t9-00}
(\psi,\phi)_{\cal PT}=\int_{C\subseteq \RR} [{\cal
PT}\psi(x)]\phi(x) dx=\int_{C\subseteq \RR} [{\cal
P}\psi(x)^*]\phi(x) dx=\int_{C\subseteq \RR} \psi(x)^*{\cal
P}\phi(x)=[\psi,\phi]_{\cal P}\, .
\ee
\item In contrast to the "usual" positive definite metric structure
of the Hilbert space $\tilde {\cal H}={\cal H}_+\oplus {\cal H}_-$,
\be{t9-0}
\left(\tilde {\cal H}, (.,.)\right), \qquad
(\psi,\phi)=\int_{C\subseteq \RR} \psi^*(x)\phi(x)d
x=\int_{C\subseteq \RR} \left(\psi^*_+\phi_+
+\psi^*_-\phi_-\right)d x\, ,
\ee
with non-negative norm
\be{t9-1}
||\psi||^2=(\psi,\psi)=||\psi_+||^2+||\psi_-||^2\ge 0\, ,
\ee
the scalar product $[\psi,\phi]_{\cal P}$ defines an indefinite
metric structure in the Krein space $\fK_{\cal P}={\cal H}_+\oplus
{\cal H}_-$, what is easily seen from the decomposition \rf{t4}
\be{t9a}
[\psi,\phi]_{\cal P}=\int_{C\subseteq \RR}
\left(\psi_+^*\phi_+-\psi_-^*\phi_-\right)dx.
\ee
\item
In rough analogy with time-like, space-like, and light-like
(isotropic) vectors in Minkowski space, one distinguishes Krein
space vectors of positive type, $[\psi_+,\psi_+]_{\cal
P}=||\psi_+||^2>0$, of negative type, $[\psi_-,\psi_-]_{\cal
P}=-||\psi_-||^2<0$, and neutral (isotropic) vectors:
\be{t9b}
[\psi,\psi]_{\cal P}=0, \quad \psi=\psi_+ +\psi_-,\quad
||\psi_+||^2=||\psi_-||^2\, .
\ee
\end{itemize}

In order to make the structural Krein space analogies of PTSQM
models and MHD dynamo setups maximally transparent, we rewrite the
eigenvalue problem, $H\psi=E\psi$, for the ${\cal PT}-$symmetric
Hamiltonian \rf{t5} in an equivalent $2\times 2$ matrix operator
representation. Introducing the projection operators
\be{t10}
P_\pm:=\frac 12 (I\pm {\cal P})
\ee
we decompose wave function $\psi$ and Hamiltonian $H$ (see, e.g.,
\cite{azizov}) as
\ba{t11}
\psi &=&P_+\psi +P_-\psi=\psi_++\psi_-\label{t10-a}\\
H&=&P_+ HP_+ + P_-HP_+ +P_+HP_- +P_- H P_- \, .\label{t10-b}
\ea
In terms of the notation
\be{t12}
H_{\pm \pm}:=P_\pm HP_\pm=-\partial^2_x+V_+(x),\qquad H_{\pm
\mp}:=P_\pm HP_\mp=iV_-(x)
\ee
this gives
\be{t13}
\left(\begin{array}{cc} H_{++} & H_{+-}\\
H_{-+} & H_{--}
\end{array}\right) \left(\begin{array}{c} \psi_+ \\
\psi_-
\end{array}\right)=E\left(\begin{array}{c} \psi_+ \\
\psi_-
\end{array}\right), \qquad {\cal P}=\left(\begin{array}{cc} I & 0\\
0 & -I
\end{array}\right)\, ,
\ee
where
\be{t14}
H_{\pm \pm}=H_{\pm \pm}^\dagger, \qquad  H_{+-}=-H_{-+}^\dagger\, .
\ee
If one replaces the matrix entries in \rf{t13}, \rf{t14} by
appropriate constants one arrives at the schematic two-level model
\be{AMeq}
 H_H
\left (\begin{array}{c}
          u \\
          v
        \end{array}
\right )
  = E\,
 \left (
 \begin{array}{c}
          u \\
          v
        \end{array}
 \right ), \qquad
 H_H=
 \left (
 \begin{array}{cc}
           c+a & b\\
 -b^* & c-a
        \end{array}
 \right ), \quad a,c \in \RR \, ,
\ee
which may be read as an elementary exemplification of Heisenberg's
linear-algebraic approach \cite{Messiah} to (${\cal
PT}-$sym\-met\-ric) Quantum Mechanics and which was intensively
studied in Refs. \cite{BBJ-1,BBJ-a,MZ-sep,most-npb-1,BMW-1,most4}.

\subsection{The spherically symmetric MHD
$\alpha^2-$dynamo\label{dynamo}}

The magnetic fields of planets, stars and galaxies are maintained by
homogeneous dynamo effects, which can be successfully described
within magnetohydrodynamics (MHD). One of the simplest dynamos is
the spherically symmetric mean-field $\alpha^2-$dynamo in its
kinematic regime. This dynamo model is capable to play a similar
paradigmatic role in MHD dynamo theory like the harmonic oscillator
in Quantum Mechanics (QM). Its operator matrix has the
form\footnote{See Appendix \ref{dynamo-physics} for a few comments
on the origin of this operator matrix and on the physics of
$\alpha^2-$dynamos.} \cite{GS-jmp1}
\be{d1}
\hat H_l[\alpha]=\left(\begin{array}{cc}-Q[1] & \alpha\\ Q[\alpha] &
-Q[1]
 \end{array}\right)
\ee
and consists of formally selfadjoint blocks
\be{d2}
Q[\alpha]:=p\alpha p + \alpha \frac{l(l+1)}{r^2}\, ,
\ee
where $p=-i(\partial_r +1/r)$  denotes the radial momentum
operator. The operator $\hat H_l[\alpha]$ is defined over an
interval $\Omega=[0,1]\ni r$ and acts on two-component vectors
$\phi$ which describe the coupled $l-$modes of the poloidal and
toroidal magnetic field components of a mean-field dynamo model
with helical turbulence function ($\alpha-$profile) $\alpha (r)$.

Although the dynamo model is not ${\cal PT}-$symmetric, its operator
$\hat H_l[\alpha]$ shares a basic underlying symmetry with PTSQM
Hamiltonians --- a $\ZZ_2-$graded pseudo-Hermiticity
($J-$pseudo-Hermiticity) \cite{most3,GSG-cz2} which is induced by
the fundamental (canonical) symmetry:
\be{d2-1} \hat H_l[\alpha]=J\hat H_l^\dagger [\alpha] J , \qquad
J=\left(\begin{array}{cc}0 & I\\ I & 0
 \end{array}\right)\, .
\ee
Similar to the reflection operator ${\cal P}$ in \rf{t6}, the
operator $J$ is unitary and involutive
\be{d3}
J^\dagger =J^{-1}, \qquad J^2=I.
\ee
The boundary conditions on the vector function $\phi$ are set at
$r=1$ (the rescaled surface radius of the star or planet whose
fluid/plasma motion maintains the dynamo effect) and it is assumed
that $\alpha(r>1)\equiv 0$. In the case of physically idealized
boundary conditions at $r=1$ (see, e.g., Ref. \cite{proctor}), the
domain ${\cal D}(\hat H_l[\alpha])$ of the operator $\hat
H_l[\alpha]$ consists of functions $\phi $ such that
\ba{d4}
{\cal D}(\hat H_l[\alpha])&:=\left\{ \phi =
\left(\begin{array}{r}\phi_1 \\ \phi_2 \end{array}\right): \
\phi\in \tilde{{\cal H}}\equiv {\cal H} \oplus {\cal H}, \
{\cal H}= L_2(\Omega,r^2 dr),\right. \nonumber\\
& \left. \Omega =[0,1], \ \phi (1)=0, \
\left.r\phi(r)\right|_{r\to 0}\to 0
 \right\},
 \ea
and $\hat H_l[\alpha]$ is self-adjoint in a Krein space
\be{d4a}
\left(\fK_J, [.,.]_J\right),\qquad
[\psi,\phi]_J=\int_0^1\psi^\dagger J\phi\, \, r^2dr,
\ee
\be{d4b}
[\hat H_l\chi,\phi]_J=[\chi,\hat H_l\phi]_J\, .
\ee
It should be noted that for physically realistic boundary conditions
\be{d5}
\left.\hat B_l\phi\right|_{r=1} =0, \qquad \hat B_l=\mbox{diag}
[\partial_r+(l+1)/r,1]
\ee
there exists no appropriate Krein space which could  make the
operator $\hat H_l[\alpha]$  $J-$self-adjoint.

The structures of PTSQM models and the $\alpha^2-$dynamo can be
compared most explicitly after passing from $\fK_J$ to an
equivalent Krein space $\fK_\mu$ with diagonal metric operator
$\mu$ and redefined  Hilbert spaces components,
$L_2(\Omega,r^2dr)\mapsto L_2(\Omega, dr)$. The diagonalization
yields
\be{d6}
J\mapsto \mu:=\left(\begin{array}{rr}I&0\\0&-I\end{array}\right)
=S^{-1}JS, \quad
S=\frac{1}{\sqrt{2}}\left(\begin{array}{rr}I&-I\\I&I\end{array}\right)\,
,
\ee
\be{d7}
\hat H_l[\alpha]\mapsto \check{H}_l[\alpha] =S^{-1}\hat
H_l[\alpha]S=\frac12\left(\begin{array}{ccc}Q[\alpha -2]+\alpha & &
-Q[\alpha] + \alpha
 \\ Q[\alpha] - \alpha &&Q[-\alpha -2]-\alpha\end{array}\right),
\ee
\be{d8}
\phi \mapsto\check{\phi}=\left(\begin{array}{c}\phi_+\\
\phi_-\end{array}\right)
=\frac{1}{\sqrt{2}}\left(\begin{array}{c}\phi_2+\phi_1\\
\phi_2-\phi_1\end{array}\right),
\ee
whereas the unitary mapping $U: \ L_2(\Omega, r^2 dr)\mapsto
L_2(\Omega, dr)$ simplifies the structure of $Q[\alpha]$ and leads
in \rf{d7}, \rf{d8} to the additional replacements
\be{d9}
\phi_{1,2}\mapsto f_{1,2}:=r\phi_{1,2},\qquad Q[\alpha]\mapsto
q[\alpha]:=rQ[\alpha]r^{-1}=-\partial_r
\alpha(r)\partial_r+\alpha(r)\frac{l(l+1)}{r^2}\, .
\ee
By inspection of \rf{t12} - \rf{t14} and \rf{d6} - \rf{d9} we find
that, in the chosen Krein space representations of the PTSQM model
and the $\alpha^2-$dynamo, the block structures of the metrics
(involution operators) ${\cal P}$ and $\mu$ coincide, ${\cal
P}=\mu$, but that the blocks of the ${\cal PT}-$symmetric
Hamiltonian and the dynamo operator show significant structural
differences\footnote{In a very rough analogy, the alpha profile
$\alpha (r)$ has some similarities to a position depending mass, as
it was studied for QM models, e.g., in Refs. \cite{var-mass}.}:
\ba{d10}
H_{\pm\pm}=-\partial_x^2+V_+(x)&\qquad \longleftrightarrow \qquad &
-q[1]\pm\frac{q[\alpha]+\alpha}{2}\, , \nn\\\nn\\
H_{\pm\mp}=iV_-(x)&\quad \longleftrightarrow \quad &\mp
\frac{q[\alpha]-\alpha}{2}\, .
\ea
It is clear that these differences in the differential expressions
(as well as the different boundary conditions on the two-component
eigenfunctions) will lead to different global behaviors of the
corresponding operator spectra. Nevertheless, both types of
systems share the same Krein-space induced features of level
crossings, what will be briefly sketched in the next subsection.

\subsection{Spectral phase transitions\label{crossings}}

Since the first PTSQM paper \cite{BB} of Bender and Boettcher it is
known that ${\cal PT}-$symmetric Hamiltonians have a real spectrum
when ${\cal PT}-$symmetry is an exact symmetry and not spontaneously
broken\footnote{We recall that this follows from \rf{i1}, \rf{t1},
the eigenvalue equation $H\psi=E\psi$ and its ${\cal
PT}-$transformed, $H{\cal PT}\psi=E^*{\cal PT}\psi$.  For real
eigenvalues, $E=E^*$, it is natural to set $\psi ={\cal PT}\psi$,
whereas $E\neq E^*$ necessarily implies $\psi \neq {\cal PT}\psi$.}
(the corresponding eigenfunctions are invariant under a ${\cal
PT}-$transformation), whereas spontaneously broken ${\cal
PT}-$symmetry is connected with complex energies. A consistent PTSQM
applicability and interpretation of the complex-energy states
remains an open question up to now (cf., e.g., \cite{kleefeld}). For
convenience, we shall call these states "unphysical" here.

\begin{figure}[htb]                     
\begin{center}                         
\epsfig{file=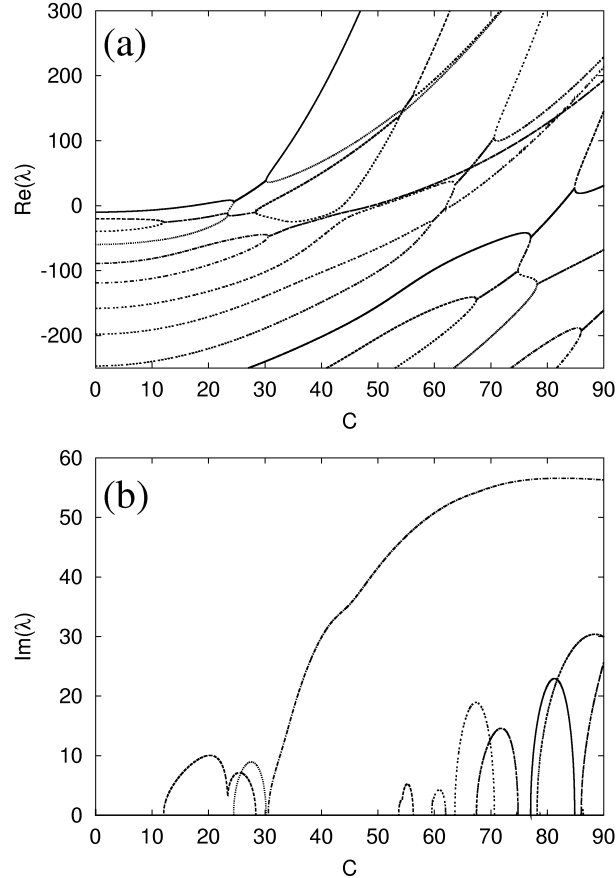,angle=0, width=0.5\textwidth}
\end{center}                         
\vspace{-2mm} \caption{Real and imaginary components of the
$\alpha^2-$dynamo spectrum as functions of the scale factor $C$ of
an $\alpha-$profile $\alpha(r)=C\times (1-26.09 \times r^2+ 53.64
\times r^3 -28.22 \times r^4)$ in the case of angular mode number
$l=1$ and physically realistic boundary conditions \rf{d5}. The
concrete coefficients in the quartic polynomial $\alpha (r)$ have
their origin in numerical simulations of the field reversal dynamics
(see Ref. \cite{SG-revers}). Only the imaginary components with $\Im
\lambda \ge 0$ are shown. The complex conjugate $(\Im \lambda \le
0)-$components are omitted for sake of brevity. \label{fig-dynamo}}
\end{figure}

The mathematically most interesting questions of PTSQM concern the
transition between the physical and unphysical domains of their
parameters. In the simple  two-state model (\ref{AMeq}) it is easy
to deduce that the quantized energies $E$ are real (``physical") for
$|a| > |b|$ while they form complex-conjugate pairs in the
``unphysical" regime where $|a| < |b|$ \cite{BBJ-a,MZ-sep}. The
boundary of its PTSQM applicability coincides with the double-cone
hypersurface in parameter space where $|a| = |b|$. One easily
verifies that whenever $|b|$ approaches $|a|$, the separate
eigen-energies $E_\pm$ as well as the corresponding two independent
bound-state eigenvectors coalesce and coincide. On the critical
hypersurface the remaining (geometrical) eigen-vector becomes
supplemented by a so called associated  vector (algebraic
eigen-vector) \cite{GSG-cz2} and the Hamiltonian matrix $H$ acquires
a Jordan-block canonical structure \cite{GSG-cz2,most-jordan}. The
latter cannot be diagonalized and it only gives the doubly
degenerate and real single ``exceptional-point" eigenvalue
$E=E_{(EP)}=c$ (cf., e.g., Ref.~\cite{Dieter-prl} for more details).

An exhaustive and consistent bound-state interpretation of the
Schr\"odinger type equation  (\ref{BBeq}) is more difficult. For
example, it requires the restriction of the range of exponents to a
finite interval of $\nu \in (-1, 2)$ for $\psi(x) \in
L_2(-\infty,\infty)$ as usual defined on the real line~\cite{BB}. A
rigorous proof of the reality of the energies turned out
unexpectedly difficult \cite{DDT-1,shin-cmp-1}. For larger exponents
$\nu$, the real line must be replaced by an appropriately  deformed
contour in the complex plane \cite{BB,BBjmp}.

A systematic analytical study of phase transition points is still
lacking for PTSQM models; the same concerns efficient mathematical
tools for deriving their location in parameter space. Similar to the
double-cone hyper-surfaces for the simple matrix model \rf{AMeq},
one expects more complicated (and more interesting) global
phase-transition hyper-surfaces in case of the Schr\"odinger type
systems. Knowing the location of these phase transition
hyper-surfaces, one would know the boundaries of the "physical"
regions of exact ${\cal PT}-$symmetry.

For the $\alpha^2-$dynamo both types of eigenvalues --- real ones as
well as pair-wise complex conjugate ones --- have a clear physical
meaning. They simply correspond to non-oscillatory and oscillatory
dynamo states, respectively. But again it is of utmost interest to
know the parameter configurations for which transitions between the
two types of states (phases) occur. In the recent paper
\cite{SG-revers}, strong numerical indications were presented that
magnetic field reversals (interchanges of North and South poles as
they are evident from paleo-magnetic data on the Earth magnetic
field \cite{reverse-book}) are induced by a special type of
nonlinear dynamics\footnote{In the concrete case, the nonlinear
transition mechanism between kinematic and saturated dynamo regime
(a brief outline of the corresponding physics can be found in
Appendix \ref{dynamo-physics}) was simulated with the help of a so
called $\alpha-$quenching (see, e.g., \cite{quench}) which simulates
the nonlinear back-reaction of the induced magnetic fields on the
$\alpha-$profile $\alpha(r)$. } in the vicinity of spectral phase
transition points.

The qualitative features of the real-to-complex phase transitions
are essentially the same for PTSQM models and for the MHD
$\alpha^2-$dynamo. They correspond to transitions from Krein space
states of positive and negative type to pair-wise neutral
(isotropic) states \cite{LT-1,GSG-cz2} --- and a square-root
branching of the spectral Riemann surface \cite{CMB-Wu,heiss-1}.
Such transitions are a generic feature of Krein-space setups and
they are new compared with setups in Hilbert spaces with purely
positive metric structures as in "usual" QM. The square-root
branching behavior is easily seen by passing from the linear
eigenvalue problems for the $2\times 2-$operator matrices $H$ and
$\hat H_l[\alpha]$ of Eqs. \rf{t13} and \rf{d1},
\be{c1}(H-E)\psi=0,\qquad
\left(\hat H_l[\alpha]-\lambda\right)\phi=0,
\ee
via substitutions
\ba{c2}\psi=\left(\begin{array}{c}
          \psi_+ \\
          -\frac{1}{H_{+-}}\left[H_{++}-E\right]\psi_+
        \end{array}
\right), \qquad \phi=\left(\begin{array}{c}
          \phi_1 \\
          \frac{1}{\alpha}\left[Q(1)+\lambda\right]\phi_1
        \end{array}
\right)
\ea
to the equivalent quadratic operator pencils
\ba{c3}\left\{(H_{--}-E)\frac{1}{H_{+-}}(H_{++}-E)-H_{-+}\right\}\psi_+&=&0,\nn\\
\left\{\left(Q[1]+\lambda\right)\frac1\alpha
\left(Q[1]+\lambda\right)-Q[\alpha] \right\}\phi_1 &=&0.
\ea
Both pencils are of the same generic operator type
\be{c4}
L[\lambda]\psi=\left[A_2\lambda^2+A_1\lambda+A_0\right]\psi=0
\ee
with a scalar product
\be{c5}
(\psi,L[\lambda]\psi)=a_2\lambda^2+a_1\lambda+a_0=0,\quad
a_j:=(\psi,A_j\psi)
\ee
which can be used to deduce the local square-root branching behavior
of the spectrum
\be{c6}
\lambda_{1,2}=\frac{1}{2a_2}\left(
-a_1\pm\sqrt{a_1^2-4a_0a_2}\right).
\ee

A typical  $\alpha^2-$dynamo spectrum with a large number of
real-to-complex transitions is presented in Fig. \ref{fig-dynamo}
(see also Refs. \cite{GSG-cz2,oscil-1}). These crossings with
real-to-complex transition occur at exceptional points (in the sense
of Kato \cite{kato}) of (square root) branching type
\cite{berry-2,heiss-2} and the corresponding eigenvalues have
geometric multiplicity one and algebraic multiplicity two
\cite{GSG-cz2}. In contrast, crossings without real-to-complex
transitions are of the same type as level crossings in Hermitian
systems \cite{LT-1} --- with geometric and algebraic multiplicity
two \cite{berry-3}. Finally, we note that although locally crossings
with real-to-complex transitions occur, in general, only between two
spectral branches, globally much more branches are involved in
mutual crossings (see Fig. \ref{fig-dynamo}). This reflects the fact
that in general the spectrum forms a multi-sheet Riemann surface
over the parameter space of the theory (see e.g.
\cite{CMB-Wu,heiss-1,Seiberg}).

In the next section, we will analyze the spectral behavior of a
${\cal PT}-$symmetric interpolation model where we will find a
similar rich structure of real-to-complex transitions as for the
$\alpha^2-$dynamo.

\section{${\cal PT}-$symmetric interpolation between square well and harmonic oscillator\label{interpol}}

In Schr\"{o}dinger-type models (\ref{BBeq}) over the open real line
$x\in (-\infty,\infty)$ a PTSQM-related separation of the
``physical" and ``unphysical" domains is, in general, a
mathematically highly non-trivial problem. Its resolution requires a
fairly subtle and rigorous mathematical argumentation
\cite{DDT-1,shin-cmp-1}. A typical result of the WKB analysis of
Ref. \cite{BB} was that in a half-open interval of $\nu \in [0, 2)$
the energies remain real and that the ${\cal PT}-$symmetry of the
wave functions remains unbroken. In parallel, a characteristic
unphysical behavior of the system (\ref{BBeq}) has been found in the
half-open interval of $\nu \in [-1, 0)$, where at any $\nu < 0$ all
the sufficiently high-lying energies $E_n$ with $n > n_0(\nu)$
``decay" in complex-conjugate pairs, $\Im E_n \neq 0$. Moreover, the
spectrum becomes empty in the Herbst-Hamiltonian limit of the
leftmost $\nu = -1$~\cite{BB}.

\subsection{Toy model ${\cal PT}-$symmetric differential
equation}

In the present section, we are going to extend the consideration of
the Schr\"odinger-type system \rf{BBeq} to exponents from the
interval $\nu \in [-2,0]$. The end-points of this interval
correspond to the purely real-valued Hermitian-system spectra of a
freely moving particle with shifted off-set energy (for $\nu=-2$)
and a harmonic oscillator (for $\nu=0$). For the exponents
$\nu\in(-2,0)$ we expect a phase of spontaneously broken ${\cal
PT}-$symmetry with an involved picture of real-to-complex spectral
phase transitions.

In order to keep the numerical analysis sufficiently simple and
robust, we assume the system located in a square
well\footnote{Various aspects of square-well-related PTSQM setups
have been earlier considered, e.g., in Refs. \cite{LT-1,sqw0,sqw}.}
(box) of {\em finite} width $2b<\infty$ and Dirichlet boundary
conditions imposed at the walls, $\psi(x=\pm b)=0$, i.e. we
introduce an IR cut-off at the low-energy end of the spectrum. This
enables us to rescale Eq. (\ref{BBeq}) to the equivalent equation
 \be{SQWeq}
 \left[ -\partial_y^2 +G\,y^2(iy)^\nu
 \right]
 \psi[x(y)] =
 \mu(E)\,
 \psi[x(y)]
  \ee
with parameter-independent boundary conditions
 \be{SQWbc}
 \psi[x(\pm 1)]=0
 \, ,
 \ee
but rescaled coupling constant and energy
 \be{int1}
 G = g\,b^{4+\nu},
 \ \ \ \ \ \
 \ \ \ \ \ \ \mu(E)=b^2\,E
 \,.
 \ee
In this notation, the original bound-state problem \rf{BBeq} with
asymptotic Dirichlet boundary conditions at $x\to \pm \infty$ is
replaced by the equivalent new problem defined within a fixed finite
interval $[-1,1]$. In the limit of very small $b \approx 0$ the
potential term becomes negligible, $G \approx 0$, and the
interaction degenerates to an infinitely deep square well (box) at
all $\nu$. A completely similar situation occurs for systems with
any non-vanishing  finite $b$, but very small exponents, $\nu
\approx -2$. In both extremal cases the problem remains exactly
solvable. The original Bender-Boettcher problem corresponds to the
strong-coupling limit, $G\to \infty$, $b\to \infty$, with $g$ hold
fixed, $g=G\, b^{-4-\nu}=\const$.

For finite coupling constants $0<G<\infty$ one expects the energy
spectrum to be divided into three sectors: into a low-energy sector
with states which are involved in real-to-complex phase transitions,
into an intermediate sector, where the $\nu-$dependent energies
still remain real, and into a high-energy sector with almost
$\nu-$independent purely real eigenvalues whose states experience
only a small perturbations from the complex interaction term. The
division into low energy and intermediate-and-high energy sectors
has been qualitatively described in a recent paper \cite{LT-1} by
Langer and Tretter who considered a square well model with an
arbitrary ${\cal PT}-$symmetric potential $V$ as perturbation.
Starting from the energy spectrum of the empty square well,
$\mu_k=k^2\pi^2/4$, \ $k=1,2,\ldots \ $, they showed that there are
no real-to-complex phase transitions for levels $k>k_s$ with $k_s$
as the lowest level satisfying the supremum bound $||V||_\infty
<(2k_s+1)\pi^2/8$. In case of our model with potential
$V(y)=gb^{4+\delta}\,y^2(iy)^\nu$ the supremum norm (see, e.g.,
\cite{RS-2}) reads (for $\nu \ge -2$)
\be{int2}
||V||_\infty=\sup_{y\in [-1,1]}|V(y)|=|V(\pm 1)|=|g|b^{4+\nu}
\ee
so that it is ensured that there are no phase transitions for
levels
\be{int3}
k>k_s(b)>\frac 12\left[\frac{8}{\pi^2}|g|b^{4+\nu}-1\right].
\ee
According to \cite{LT-1} it holds for the corresponding real
eigenvalues $\mu_k$: $|\mu_k-k^2\pi^2/4|<|g|b^{4+\nu}$.  The
supremum bound is safe, but at the same time rather rough
\cite{CT-private}. The subsequent exact numerical analysis shows
that, depending on the concrete exponents $\nu$, the real-to-complex
phase transitions  in the model \rf{SQWeq} stop at much lower energy
levels.

\subsection{The emergence of $ \Im E \neq 0$ on certain
finite subintervals of $\nu \in (-2,0)$}

In the generic case with $\nu \in (-2,0)$ and $b > 0$, we have
solved Eqs. (\ref{SQWeq}) + (\ref{SQWbc}) numerically by means of a
shooting technique with a fifth-order Runge-Kutta method, utilizing
and adapting standard routines from Numerical Recipes \cite{RECIP}.
The corresponding code had been validated extensively in earlier
work by comparison with known analytical results and other numerical
results in dynamo theory and quantum mechanics.
\begin{figure}[thb]                     
\begin{center}                         
\epsfig{file=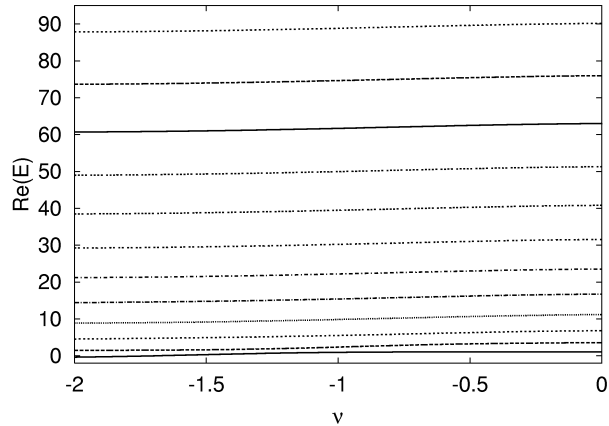,angle=0, width=0.5\textwidth}
\end{center}                         
\vspace{-2mm} \caption{Spectrum of the ${\cal PT}-$symmetric
interpolation Hamiltonian $H=-\partial^2_x+x^2\left(ix\right)^\nu$
as function of the exponent $\nu$ for the cut-off length $b=2$. All
eigenvalues are real and almost independent of $\nu$. The spectrum
is only slightly deviating from that of an empty square
well.\label{figb2}}
\end{figure}
\begin{figure}[thb]                     
\begin{center}                         
\epsfig{file=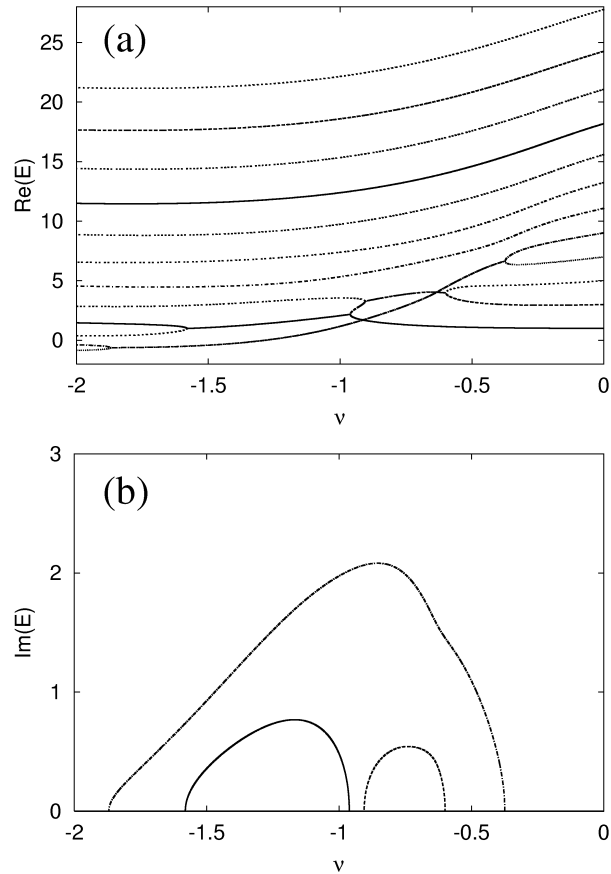,angle=0, width=0.5\textwidth}
\end{center}                         
\vspace{-2mm} \caption{Real and imaginary components of the spectrum
in the case of a cut-off length $b=4$ (complex conjugate $(\Im
\lambda \le 0)-$components omitted, as well as further higher lying
levels without real-to-complex transitions). The low-energy sector
with its multiple real-to-complex transitions starts to form.
\label{figb4}}
\end{figure}
\begin{figure}[thb]                     
\begin{center}                         
\epsfig{file=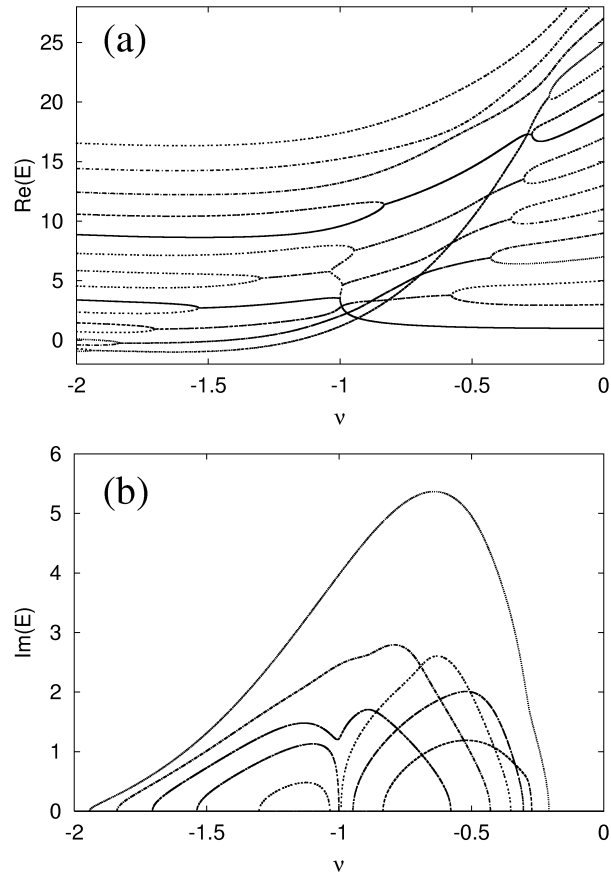,angle=0, width=0.5\textwidth}
\end{center}                         
\vspace{-2mm} \caption{Spectrum for a cut-off length $b=6$.
\label{figb6}}
\end{figure}
\begin{figure}[thb]                     
\begin{center}                         
\epsfig{file=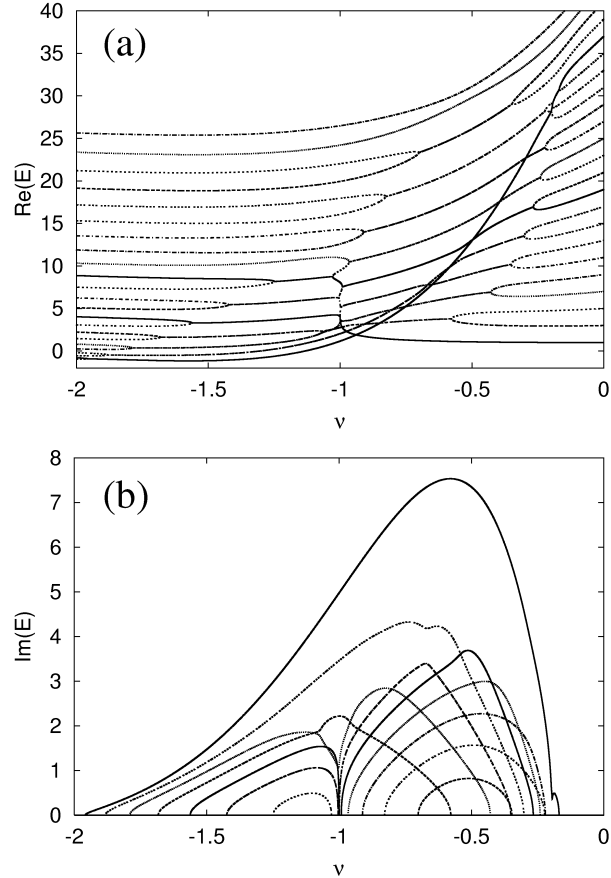,angle=0, width=0.5\textwidth}
\end{center}                         
\vspace{-2mm} \caption{At a cut-off length $b=7$ the generic
structure of the spectrum is clearly visible. The web-like pattern
of the real components (Fig. \ref{figb7}a) contains purely real
branches in the vicinity of the left $(\nu=-2)$ and right $(\nu=0)$
end points of the considered interval as well as a chain of purely
real intermediate segments in the vicinity of $\nu=-1$. (See Fig.
\ref{figb7-detail} for a detailed view.) The increasing number of
imaginary components (Fig. \ref{figb7}b) with high gradients
$|\partial_\nu E(\nu\approx -1)|\gg1$ which accumulate in the
vicinity of $\nu=-1$ are first indications of the formation of a
local spectral singularity at $\nu=-1$.\label{figb7}}
\end{figure}
\begin{figure}[thb]                     
\begin{center}                         
\epsfig{file=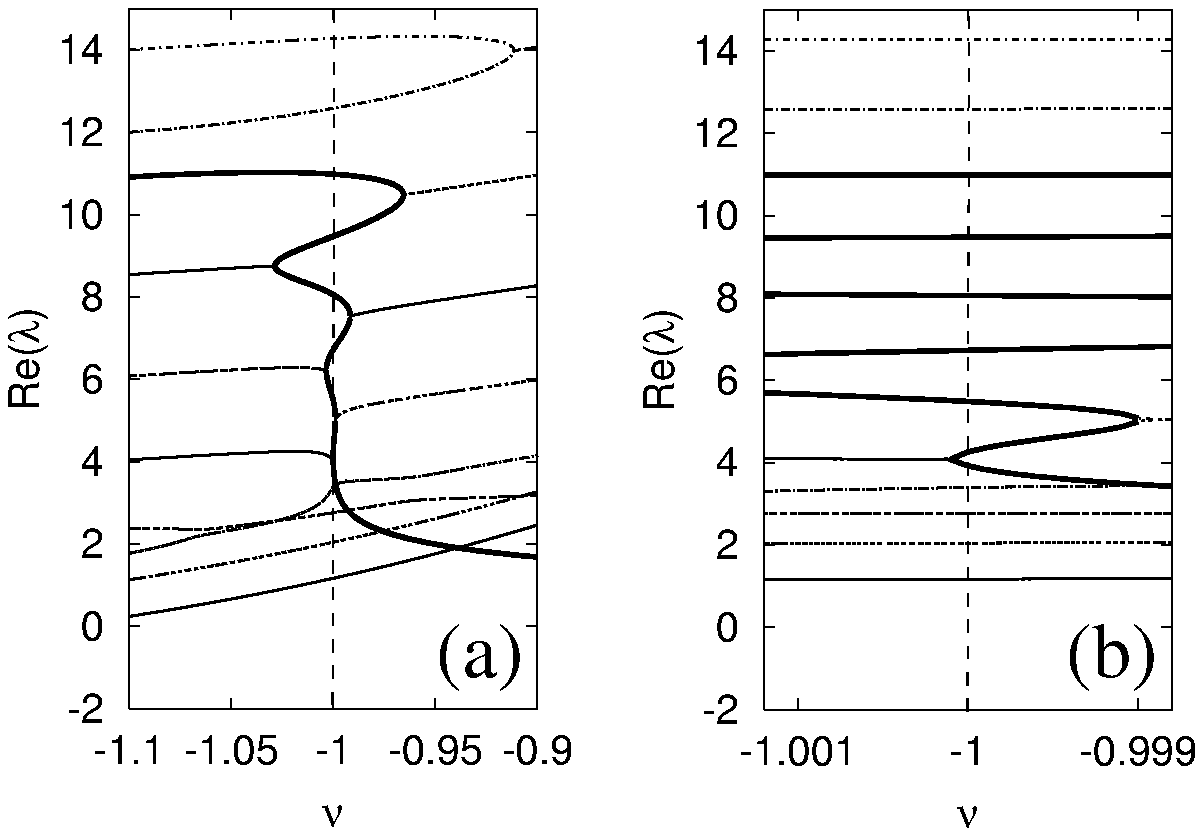,angle=0, width=0.6\textwidth}
\end{center}                         
\vspace{-2mm} \caption{The purely real curve (high-lighted fat) in
the vicinity of $\nu=-1$ (cut-off length $b=7$) is formed by
intermediate real segments between complex valued segments which
branch off to the left and to the right. This leads to a ladder-like
structure with exceptional points as nodes. The zoomed picture in
Fig. \ref{figb7-detail}b shows that the "oscillations" of the real
curve about the line $\nu=-1$ persist also at its lower end, but
with strongly reduced "amplitude". When the cut-off is slightly
increased to $b\gtrsim 7$ the lowest exceptional (real-to-complex
transition) point (Fig. \ref{figb7-detail}b) will cross the line
$\nu=-1$ and will in the region $\nu>-1$ coalesce with the other
(nearest) exceptional point. As result one intermediate segment will
be removed from the real curve and a purely complex-valued
branch-pair will smoothly tend from "far left" $(\nu<-1)$ to "far
right" $(\nu>-1)$ --- similar to the lower lying purely complex
branches visible in the graphics.\label{figb7-detail}}
\end{figure}

A sample of the results of such a study is depicted in Figures
\ref{figb2} -- \ref{figb7-detail}, where we have chosen $g=1$ and
displayed the first few energy levels $E(\nu)$ over the entire
interval $\nu \in (-2,0)$ for the sequence of values $b = 2, 4, 6,
7$. The important results of this numerical experiment are the
following:

\begin{itemize}

\item
At all sufficiently small $b$, as sampled in Fig. \ref{figb2}, the
energy spectrum  exhibits a more or less $\nu-$independent
square-well form.

\item
At not too large $b$ the spectrum, as sampled in Fig. \ref{figb4},
proves clearly separated into the high-lying part (where the
energies still preserve their approximate $\nu-$independence), an
intermediate perturbative part (where the perceivably
$\nu-$dependent energies still remain all real) and the low-lying
part (where one encounters the first real-to-complex phase
transitions).

\item The actual lowest level numbers $k_c(b)$ (critical level numbers)
of the modes which are not involved in real-to-complex transitions
lie much below the safe supremum bounds $k_s(b)$ of inequality
\rf{int3}. Choosing, for example, the exponents $\nu=-1/2$ and
$\nu=-3/2$ we read off that
\ba{int4}\nu=-1/2:&\qquad & b=\left\{\begin{array}{c}
  2 \\
  4 \\
  6 \\
  7
\end{array}\right.,\quad k_s(b)>\left\{\begin{array}{c}
  4.08 \\
  51.4 \\
  213 \\
  367
\end{array}\right.,\quad k_c(b)=\left\{\begin{array}{c}
  1 \\
  6 \\
  14 \\
  22
\end{array}\right.,\nn\\
\nu=-3/2:&\qquad & b=\left\{\begin{array}{c}
  2 \\
  4 \\
  6 \\
  7
\end{array}\right.,\quad k_s(b)>\left\{\begin{array}{c}
  1.79 \\
  12.5 \\
  35.2 \\
  52.1
\end{array}\right.,\quad k_c(b)=\left\{\begin{array}{c}
  1 \\
  5 \\
  9 \\
  11
\end{array}\right.\, .
\ea

\item
Starting from the ``intermediate width" region, sampled at $b = 6$
in Fig. \ref{figb6}, we find that the {\em left-hand half} of the
picture exhibits a clear transition from the slightly non-Hermitian
square well regime (with the higher energies all real) to its more
strongly non-Hermitian extension where for all exponents $\nu$ not
too distant from $\nu=-2$ the purely imaginary and finite component
of the potential resembles the spatially antisymmetric part of the
exactly solvable ${\cal PT}-$symmetric Heavyside step potential
within a square well considered in Ref. \cite{sqw0}. This explains
why in Figs. \ref{figb4} - \ref{figb7} the continuing decrease of
$\nu$ makes the respective two or three lowest pairs of the energies
merge and complexify.

\item
At ``sufficiently large" cut-offs $b$, all the {\em real} low-lying
energies depicted in the {\em right-hand halves} of Figs.
\ref{figb6} and \ref{figb7} obviously stabilize and approach the
$b\to \infty$ limiting pattern as published in Ref. \cite{BB,DDT-1}.
In particular, we see that the ground-state energy remains real and
that it starts growing more quickly only when the values of $\nu$
move down and closer to the Herbst limit of $\nu\to -1^+$. We
observe that in a more appropriate way this growing real branch
should be interpreted as a special type of ladder-shaped merger of
intermediate real segments which actually correspond to level-pairs
with higher mode numbers. A zoomed view on this peculiarity is
presented in Fig. \ref{figb7-detail}, where it is clearly visible
that a chain of exceptional points is located on this branch with
alternating complex-valued segments branching off to the left and to
the right. These segments fit, after further complex-to-real
transitions, to the real eigenvalues of the $\nu\to 0$ and $\nu\to
-2$ limit models.

\item When the cut-off $b$ is increased the following simultaneous
changes in the spectrum can be observed. In the upper low-energy
region with $\nu >-1$ step by step more and more level pairs become
twisted into the complex sector. With a "slight delay in $b$" and at
$\nu <-1$ the lower of the twisted levels undergo a second pairwise
real-to-complex transition with the levels below them. A sort of web
structure is forming with a purely real branch remaining between the
left $(\nu<-1)$ and right $(\nu >-1 )$ purely complex (twisted)
spectral regions. The complex-valued level pairs are branching off
from the real branch forming a ladder-shaped structure. At the
low-energy end of this ladder a second process occurs. The left
$(\nu<-1)$ complex level pairs are passing the line $\nu=-1$ and
move to the right of it. There, at some $\nu>-1$ the corresponding
exceptional point merges with an exceptional point of a right
branch. As result one of the real segments between left and right
off-branching levels disappears and a smooth complex-valued branch
forms which extends over a large $\nu$ interval and whose imaginary
components are increasing very fast when $b$ is increased. It
remains the real branch which becomes more and more vertical whereas
the complex branch is not intersecting with it (the real component
of the complex branch is coinciding at one point with the real
branch but the imaginary components are not coinciding).

\end{itemize}
Analyzing the sequence of Figs. \ref{figb2} - \ref{figb7-detail} we
observe that, when the cut-off $b$ is increased, a rather special
(and seemingly inextricable) branch pattern\footnote{The phenomenon
may be generic since in Ref. \cite{MZ-wiggly}, the "wiggly upwards"
spectral pattern has also been detected for a very different
one-parametric family of asymptotically exponential ${\cal
PT}-$symmetric potentials $V(x)=-\left(i \sinh x\right)^\beta$ near
the Herbst-like exponent $\beta=1$.} of real and complex eigenvalues
is forming in the vicinity of the exponent $\nu=-1$. The extreme
steepness of an increasing number of imaginary branches and their
accumulation at $\nu=-1$ (see Figs. \ref{figb6}, \ref{figb7}) as
well as the occurrence of the almost vertical real branch (Fig.
\ref{figb7-detail}) are indicating the formation of a local spectral
singularity with $\left.\partial_\nu E\right|_{\nu \to -1}\to \pm
\infty$ at the (almost) vertical segments of the real-valued branch
as well as on the imaginary branches close to the exceptional points
of the "ladder" structure. From the figures it is not at all obvious
how this pattern is compatible with the Herbst limit of an empty
spectrum for $b\to \infty$ at $\nu=-1$. We will resolve this
interesting puzzle in the next section.

\section{The Herbst limit and its relation to the Squire equation of hydrodynamics\label{Herbst-box}}

In Ref. \cite{Herbst-1} it was was shown by I. Herbst that the
spectrum of a Hamiltonian \rf{h1} with imaginary linear potential
(imaginary homogeneous electric field) over the real line $x\in \RR$
is empty. The differential expression of the corresponding operator
coincides with that of  the ${\cal PT}-$symmetric Schr\"odinger type
equation \rf{SQWeq} with exponent $\nu=-1$,
\be{h1}
\left[ -\partial_y^2 -igb^3\,y
 \right]
 \psi(y) =
 b^2E\,
 \psi(y),\qquad \psi(y=\pm 1)=0.
\ee
The only difference of \rf{h1} compared to the Herbst model is in
the Dirichlet boundary conditions at $y=\pm 1$ which restrict the
system to a box (square well). Due to this analogy and for sake of
brevity, we will call the model \rf{h1} a "Herbst box".

We start our consideration by noticing that Eq. \rf{h1} and the
spectral function $\mu (b,E)=b^2E$ are invariant under the
rescaling $b\mapsto g^{-1/3}b, \ E\mapsto g^{2/3}E$ so that
henceforth we can set $g=1$, without loss of generality. The
corresponding Herbst box Hamiltonian we denote as
\be{h1-0}
H_{Hb}(b):=-\partial_y^2 -ib^3\,y\, .
\ee
Equation \rf{h1} itself is of Airy type and its solutions can be
expressed as
\ba{h2}
\psi(y)&=&C_1 A_1[\xi(y)]+C_2 A_2[\xi(y)],\label{h2a}\\
\xi(y)&:=& e^{i\frac \pi 3}\left[-iby-E\right],\label{h2b}
\ea
where $C_{1,2}=\const$, and $A_1(\xi)$, $A_2(\xi)$ are any two of
the Airy functions $\Ai (\xi), \ \Ai (q\xi), \ \Ai (q^2\xi)$ with
$q:=e^{i2\pi/3}$. As usual, the boundary conditions lead to a
characteristic determinant which defines the spectrum of the
eigenvalue problem. In case of Eq. \rf{h1}, it reads
\be{h3}
\Delta(E)=A_1 (\xi_+)A_2 (\xi_-)-A_1 (\xi_-)A_2 (\xi_+)=0, \qquad
\xi_\pm:=\xi(y=\pm 1).
\ee
Characteristic determinants of this type (built over Airy functions)
have been intensively studied since 1995 in a paper series of Stepin
\cite{Step-1,Step-2} and Shkalikov \etal
\cite{Shkalikov}\footnote{For related work see also Ref.
\cite{Red-1}.} on the spectral properties of the Squire equation of
hydrodynamics\footnote{The corresponding physical background can be
found, e.g., in \cite{RSH,SH}.}
\be{h4}
H_{Sq}(\varepsilon):=i\varepsilon\partial_y^2 +y, \qquad
\left(H_{Sq}-\lambda\right)\chi=0,\qquad \chi(y=\pm1)=0,\qquad
\varepsilon:=\left(\tilde\alpha R\right)^{-1}.
\ee
Before we make the (obviously existing) relation of this model to
the Herbst box model explicit, we briefly review a few of its
properties.

The Squire equation \rf{h4} describes the normal vorticity of a
plane Couette flow with linear velocity profile. The parameter
$\tilde \alpha>0$ denotes a real-valued wave number which originates
from the decomposition of a 2D-flow-perturbation
\be{s2}
\Psi(x,y,t)=\chi(y)e^{i\tilde \alpha (x-\lambda t)},
\ee
$R>0$ is the Reynolds number and $\varepsilon$ --- the viscosity.
The spectrum of $H_{Sq}$ was found to have a $Y-$shaped form
\cite{Step-2,Shkalikov,RSH}. All the eigenvalues are located in a
close vicinity of the three segments $(1, -i/\sqrt 3]$, $(-1,
-i/\sqrt 3]$, $[-i/\sqrt 3, -i\infty)$. In the limit of large $R \to
\infty$, $\tilde \alpha \ge 1$ and correspondingly small
$\varepsilon \to 0^+$ the eigenvalue problem \rf{h4} turns into a
singular perturbation problem and its eigenvalues show a remarkable
limiting behavior: For $\varepsilon\to 0^+$, more and more
eigenvalues "move in" from $-i\infty$ along the line $[-i/\sqrt 3,
-i\infty)$, merge pairwise in the vicinity of the point $-i/\sqrt 3$
and depart then (again pairwise) to move symmetrically along the
segments $[1, -i/\sqrt 3]$, $[-1, -i/\sqrt 3]$ and to "fill" them
step by step --- leaving the $Y$ shape-invariant. The process was
described in Ref. \cite{Step-2} as a special type of transition from
a discrete spectrum to a continuous one. Explicitly, the following
asymptotic estimates were found in \cite{Step-2,Shkalikov}
\ba{h5}\lambda_n&\sim & -i\varepsilon \frac{\pi^2 n^2}{4}\  \in \ [-i/\sqrt 3, -i\infty),\quad
n\to \infty \label{h5a}\\
\lambda_n^\pm &\sim & \pm 1\pm \varepsilon^{1/3}s_n \, e^{\pm
i\pi/6}\ \in \ (\pm 1, -i/\sqrt 3],\quad \varepsilon\to 0^+
\label{h5b},
\ea
where $s_n$ are the zeros of the Airy function
\be{h6}
\Ai (s_n)=0,\qquad s_n\in \RR_-\, .
\ee
One clearly sees that the smaller $\varepsilon$ is chosen the
smaller the distances between the eigenvalues become --- leading in
the limit $\varepsilon\to 0$ to a quasi-continuous spectrum.

Noticing that the pairwise merging and splitting (level crossing) of
the eigenvalues occurs at $\lambda^+\sim \lambda^-\sim -i/\sqrt 3$,
it is easy to estimate that the value $\varepsilon_n$, for which
this crossing is connected with the $n-$th Airy function root $s_n$,
is given by
\be{h6-1}
\varepsilon_n^{1/3}\sim \frac{2}{|s_n|\sqrt 3} \ .
\ee

Let us utilize the above results now for the Herbst box model. A
simple comparison of the eigenvalue problems \rf{h1} (for $g=1$) and
\rf{h4} shows that these problems may be made coinciding if one sets
\be{h7}
i\varepsilon^{-1}\left[H_{Sq}(\varepsilon)-\lambda\right]\chi(y)\stackrel{!}{=}{\cal
P}\left[H_{Hb}(b)-b^2E\right]{\cal P}\left[{\cal P}\psi(y)\right]=0
\ee
and identifies
\be{h8} b^3=\varepsilon^{-1},\qquad
E=ib\lambda,\qquad
 {\cal
P}H_{Hb}(b){\cal P}=i\varepsilon^{-1} H_{Sq}(\varepsilon),\qquad
{\cal P}\psi (y)=\chi (y)\, .
\ee
This means that the two models are related by the combined action of
a rescaling, a Wick-rotation and a coordinate reflection ${\cal P}$.

With the help of the estimates \rf{h5a}, \rf{h5b} it is now an easy
task to explain the behavior of the Herbst-box spectrum $E(b)$.
\begin{itemize}
\item The rescaled spectrum $E(b)/b=i\lambda(\varepsilon=b^{-3})$
(shown in Fig. \ref{figY}) is simply the Wick-rotated version of the
original shape-invariant "Y" of the Squire operator
$H_{Sq}(\varepsilon)$.
\begin{figure}[thb]                     
\begin{center}                         
\epsfig{file=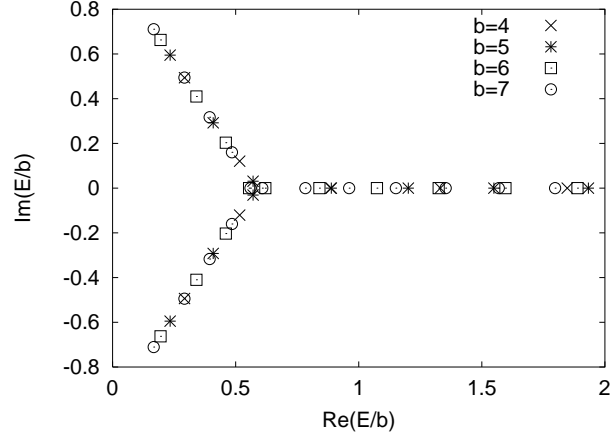,angle=0, width=0.5\textwidth}
\end{center}                         
\vspace{-2mm} \caption{The rescaled Herbst-box spectrum
$E/b=i\lambda(\epsilon=b^{-3})$ coincides with the Wick-rotated
$Y-$shaped spectrum of the Squire operator
$H_{Sq}(\varepsilon)$.\label{figY}}
\end{figure}
With increasing $b=\varepsilon^{-1/3}$ more and more eigenvalues
"move in" from $+\infty$ and "fill" the two complex conjugate
branches $(i,1/\sqrt 3]$, $(-i,1/\sqrt 3]$ of the "Y" as well as the
half-line $[1/\sqrt 3,+\infty)$ --- in a similar way as in the
original $\lambda(\varepsilon\to 0)$ limit. For $b\to \infty$ the
spectrum becomes quasi-continuous on the rotated "Y".

\item Due to the shape invariance of $E(b)/b$, the spectrum $E(b)=ib\lambda(b^{-3})$ itself inflates
when $b$ increases. It is located in the close vicinity of the
segments $(ib, b/\sqrt 3]$, $(-ib, b/\sqrt 3]$, $[b/\sqrt 3,
+\infty)$ and moves with $b\to \infty$ to infinity --- leaving (for
sufficiently high $b$) an empty region at any fixed finite distance
from the origin of the spectral $E-$plane. Hence, we find (as
required) that for $b\to \infty$ the Herbst box spectrum turns into
the empty spectrum of the original Herbst model over the real line
$\RR$.

\item For finite $b$, the asymptotic estimates \rf{h5a}, \rf{h5b} map
into
\ba{h9}E_k&\sim &  \frac{\pi^2 k^2}{4b^2}\  \in \ b\times [1/\sqrt 3, +\infty),\quad
k\to \infty \, ,\label{h9a}\\
E_n^\pm &\sim & \pm ib\pm i s_n \, e^{\pm i\pi/6}\ \in \ b\times(\pm
i, 1/\sqrt 3\, ],\quad b\to \infty \label{h9b}\, ,
\ea
and we identify \rf{h9a} as the pure square well spectrum
\be{h10}
\mu(b,E_k)=b^2\, E_k\sim \pi^2 k^2/4 \ \in \ b^3\times [1/\sqrt 3,
+\infty)
\ee
of the high-energy sector which is almost not affected by the
${\cal PT}-$symmetric interaction.
\begin{figure}[htb]                     
\begin{center}                         
\epsfig{file=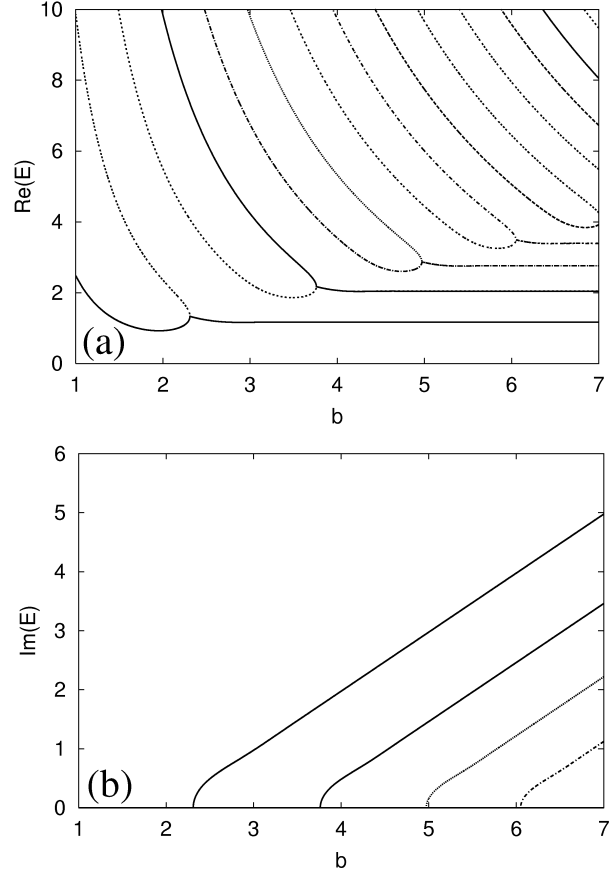,angle=0, width=0.5\textwidth}
\end{center}                         
\vspace{-2mm} \caption{Real and imaginary components of the
Herbst-box spectrum as functions of the cut-off length $b$ (complex
conjugate $(\Im \lambda \le 0)-$components omitted). The
asymptotical behavior of the complex-valued branches is clearly
visible (constant real components and linear $b-$dependence of the
imaginary components).\label{figE-b-1}}
\end{figure}
In contrast, the low-energy sector described by \rf{h9b} shows a
purely linear scaling behavior of the imaginary energy components
$\Im E_n^\pm$, whereas the real components remain asymptotically
fixed when $b$ increases. This situation is also clearly visible
from the numerical results presented in Fig. \ref{figE-b-1}.
\begin{figure}[htb]                     
\begin{center}                         
\epsfig{file=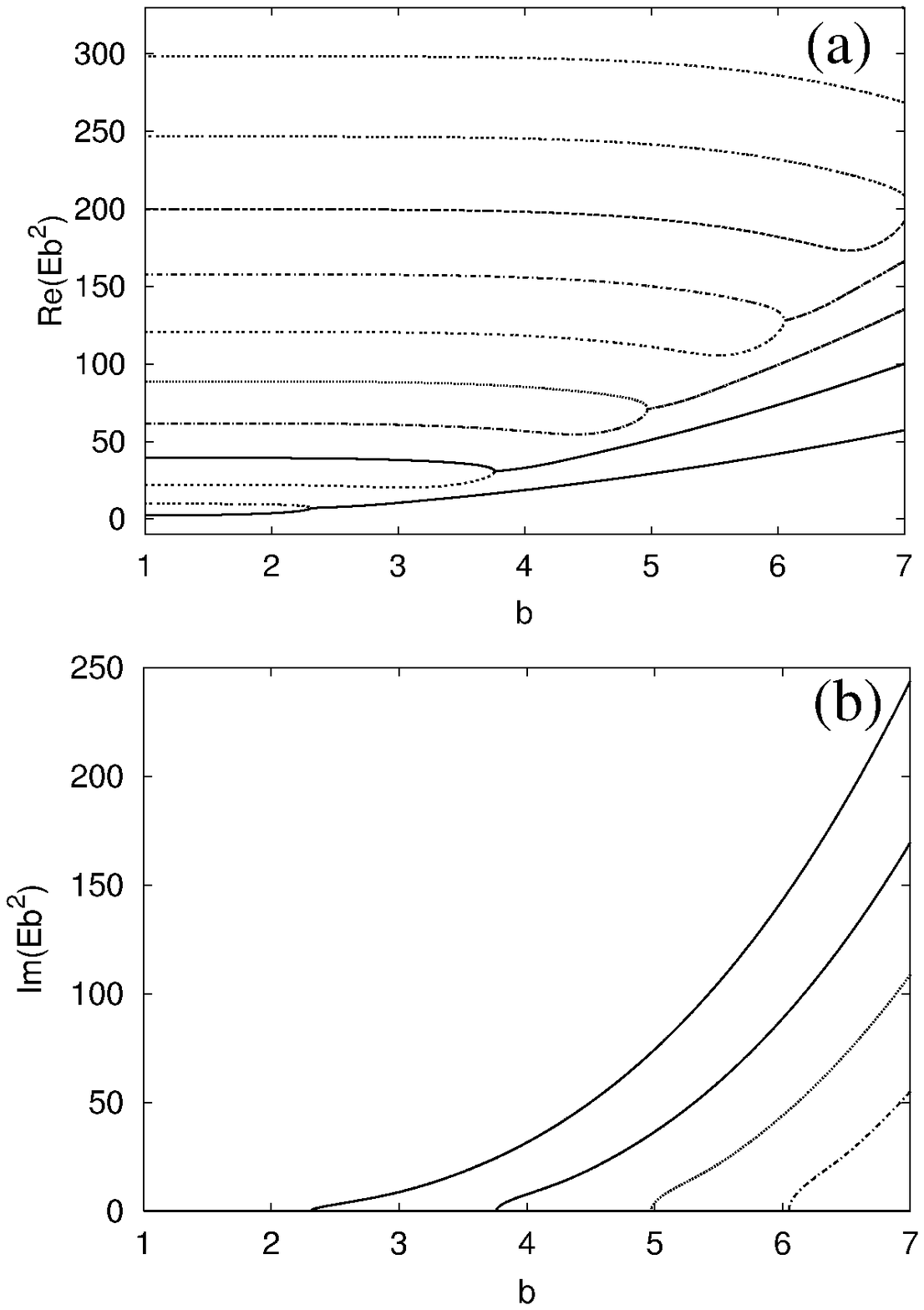,angle=0, width=0.5\textwidth}
\end{center}                         
\vspace{-2mm} \caption{The rescaled Herbst-box spectrum
$\mu(b,E)=b^2 E$ allows a complementary view on the transition from
the high-energy sector to the intermediate and low-energy sector.
\label{figE-b3-1}}
\end{figure}
The graphics of the equivalent spectrum $\mu(b,E)$, depicted in Fig.
\ref{figE-b3-1}, provides a complementary description and shows how
(for increasing $b$) the $b-$independent eigenvalues $\mu(b,E_k)$ of
Eq. \rf{h10} leave the high-energy sector, obtain an explicit
$b-$dependence in the intermediate-energy sector and finally
coalesce and split into complex conjugate pairs.

\item From the form of the spectral branches on the $E-$plane (rotated
"Y") it is clear that the level-crossings in the vicinity of
$E\sim b/\sqrt 3$ correspond to the typical real-to-complex phase
transitions of ${\cal PT}-$symmetric models in Krein spaces. With
the help of relation \rf{h6-1} the cut-off-scales $b_n$ and
positions $E_n$ of the level crossings  can be roughly estimated
as
\be{h11}
b_n\sim |s_n|\sqrt 3 /2, \qquad E_n\sim |s_n|/2\, .
\ee
One can use the explicit values of these $b_n$, $(b_1\approx
2.02,b_2\approx 3.54, b_3\approx 4.78, b_4\approx 5.88, b_5\approx
6.87, b_6\approx 7.81,\ldots)$, to roughly derive the number of the
lowest uncrossed modes in the cases $b=2,4,6,7$. For $b=2,4,6$ the
result exactly coincides with the level crossing pattern shown (at
$\nu=-1$) in Figs. \ref{figb2} - \ref{figb6}, whereas the value
$b_5$ is clearly smaller than the actual transition value $b_{5(c)}$
for which according to Figs. \ref{figb7-detail}b and \ref{figE-b-1}
holds $b_{5(c)}\gtrsim 7$.

For completeness, we note that the asymptotic approximation
\cite{Step-2} of the Airy function roots
\be{h11-1}
|s_n|=\left[\frac{3\pi}{2}\left(n-\frac
14\right)\right]^{2/3}+O(n^{-4/3}), \qquad n\to \infty
\ee
together with \rf{h11} yields the following rough estimate for the
lowest purely real-valued mode
\be{h11-2}
k_a>\frac{4}{3\pi}\left(\frac{2}{\sqrt 3}\; b \right)^{3/2}+\frac
12\, .
\ee
The scaling dimension $\kappa_a=3/2$ of this bound is only one half
of the scaling dimension $\kappa_s=4+\nu=3$ of the corresponding
supremum bound \rf{int3}.

The exact positions of the level crossing points are given by the
multiple roots of the characteristic determinant, $\Delta(E)=0$, \
$\partial_E\Delta(E)=0$. This equation system can be simplified via
Wronskian $W[A_1(.),A_2(.)]$ to yield  the conditions
\be{h12} A_{1,2}(\xi_+)=\pm A_{1,2}(\xi_-)
\ee
(see \cite{Step-2} for the details).

Plugging the numerical results from the eigenvalue solver into this
equation with $A_{1,2}(\xi)$ chosen as in \cite{Step-2},
$A_1(\xi):=\Ai (\xi)$, $A_2(\xi):=\Ai(q^2\xi)$, selects the
condition $A_{1,2}(\xi_+)=A_{1,2}(\xi_-)$ and satisfies it within
numerical working precision. For the same data holds
$A_{1,2}(\xi_+)\neq -A_{1,2}(\xi_-)$.

\item The spectral behavior $E(b)$ for increasing cut-off $b$ can be
summarized as follows. At the beginning, the real eigenvalues from
the high-energy sector decrease as $E_k\sim \pi^2 k^2/(4b^2)$
--- moving into the intermediate energy region. When $b$
approaches $b_n\sim |s_n|\sqrt 3 /2$ from below, the real
eigenvalues $\{E_{2n-1},E_{2n}\}$ \ (corresponding to a pair of
positive and negative Krein space states \cite{LT-1}) coalesce at
$E_{2n-1}\sim E_{2n}\sim |s_n|/2$ and a real-to-complex transition
occurs $\{E_{2n-1},E_{2n}\}\longrightarrow \{E_n^+,E_n^-\}$. When
$b$ is further increased the real energy components remain fixed
$\Re E_n^\pm\approx |s_n|/2$ (see Fig. \ref{figE-b-1}a), whereas the
imaginary components blow up linearly along the asymptotes $\Im
E_n^\pm\sim \pm b\pm s_n\sqrt 3/2$ (Fig. \ref{figE-b-1}b).

\end{itemize}

Let us, for finite $b$, relate the obtained Herbst-box results to
the spectral behavior of the ${\cal PT}-$symmetric interpolation
model of the previous section. Apart from the obvious one-to-one
correspondence of the high-energy sectors (see \rf{h10}), a clear
identification is immediately possible for those Herbst-box
eigenvalues which are close to the imaginary axis and which have the
largest imaginary components. These eigenvalues are located on the
branches with the largest imaginary components in Figs. \ref{figb4}
- \ref{figb7} (which stay complex when $\nu$ passes through the
Herbst-box value $\nu=-1$). It is clearly visible from these figures
that, for increasing $b$, the imaginary components are blowing up,
whereas the real components remain asymptotically constant.

\begin{figure}[hbt]                     
\begin{center}                         
\epsfig{file=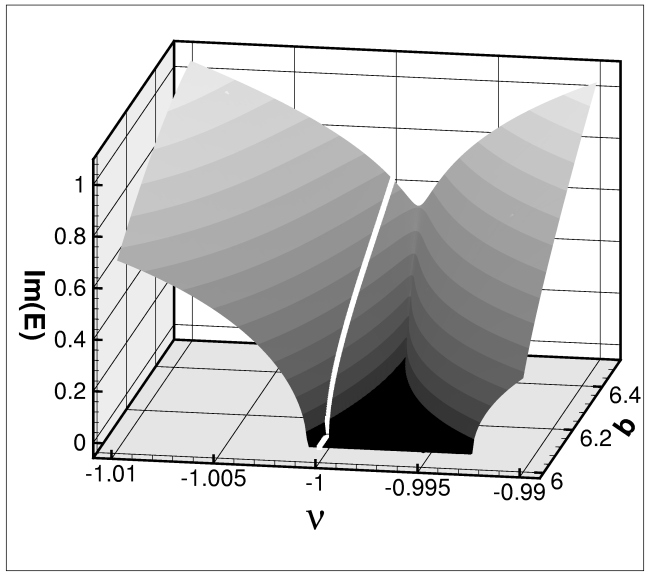,angle=0, width=0.5\textwidth}
\end{center}                         
\vspace{-2mm} \caption{Concrete example for the generic merging
process of two complex-valued spectral branches (present here for
$b< b_{coal}\approx 6.36$) into a single complex-valued branch for
$b>b_{coal}$. The two exceptional (real-to-complex transition)
points existing for $b<b_{coal}$ are located on the plane
$(\nu,b,\Im E=0)$ and coalesce at $b=b_{coal}$. The corresponding
point $(\nu_{coal}\approx-0.9983
>-1,b_{coal}\approx 6.36,\Im E=0)$ is the starting (cusp) point of a
sharp "valley" of the non-vanishing imaginary component which
steeply grows and smooths when $b>b_{coal}$ is further increased.
The white curve marks the Herbst-box values. Before the two
exceptional points coalesce, the left one of these points crosses
the Herbst-box configuration at $(\nu=-1,b=b_c\approx 6.02,\Im E=0)$
and a real-to-complex transition occurs for the Herbst-box model.
\label{fig-riemann}}
\end{figure}

So far, we have found a clear correspondence for  those regions on
the Herbst-box "Y" which are located away from the center of the "Y"
with its real-to-complex phase transitions. A more subtle situation
occurs in the vicinity of this center. The corresponding Herbst box
eigenvalues will map into points located close to (or on) the
forming (almost) vertical segment of the purely real branch depicted
in Fig. \ref{figb7-detail}. From the zoomed graphics in Fig.
\ref{figb7-detail}b we observe that the purely real and almost
vertical branch of the interpolation model "oscillates" around the
Herbst box line at $\nu=-1$ with strongly decreasing "amplitude" to
its low-energy part. With increasing $b$ this "decreasing amplitude"
effect becomes stronger and the "oscillations" are only traceable
with the help of an appropriately increased zooming scale.
Nevertheless, it can be read off that the real-to-complex transition
of the Herbst box spectrum follows qualitatively the same scheme for
any finite $b$. The real eigenvalues of the Herbst box are all
located on the purely real branch of the interpolation model and the
real-to-complex transition occurs when the lowest exceptional point
on this branch moves from the left sector $\nu<-1$ through the
Herbst-box value $\nu=-1$ into the right sector $\nu>-1$ --- to
coalesce afterwards with the next higher exceptional point from the
right sector. With this passing of the left-sector exceptional point
through the line $\nu=-1$ the Herbst box eigenvalues become pairwise
complex conjugate with strongly increasing imaginary components (due
to the asymptotically diverging gradient $\left.\partial_\nu E(b\to
\infty)\right|_{\nu\approx -1}\longrightarrow\pm \infty$). The
real-to-complex transition with subsequently increasing imaginary
components are illustrated in Fig. \ref{fig-riemann}.

Finally, we note that in the limit $b\to \infty$, the lowest-lying
intersection of the purely real branch with the Herbst-box line
$\nu=-1$ moves away to infinity like $b/\sqrt 3$ (the lower bound of
the real segment $[b/\sqrt 3, \infty)$ of the Herbst-box "Y") so
that the real branch itself remains for any finite energy in the
right sector $\nu>-1$ --- approaching the Herbst-box line
asymptotically. This reproduces the earlier observations of Refs.
\cite{BB,DDT-1} for the spectrum of the Bender-Boettcher problem
over the real line. Additionally, our Herbst box results predict for
this problem diverging imaginary components at $\nu=-1$: $|\Im
E(b\to \infty,\nu\to -1)|\to \infty$. Taking these observations
together we once more see that in the limit $b\to \infty$ a spectral
singularity is forming at $\nu=-1$ with $|E(b\to\infty,\nu\to
-1)|\to \infty$, $|\partial_\nu E(b\to\infty,\nu\to -1)|\to \infty$.

\section{Conlusions\label{conclu}}

In the present paper we considered three models emerging in
different physical setups, but which are closely related with each
other by their underlying mathematical structure as spectral
problems in Krein spaces. The models are a one-dimensional ${\cal
PT}-$symmetric quantum mechanical interpolation setup defined over a
square well of finite width $2b$, the spherically symmetric MHD
$\alpha^2-$dynamo as well as the Squire equation of hydrodynamics.
For the PTSQM model and the $\alpha^2-$dynamo we made their close
relation transparent by transforming them into a $2\times 2$ matrix
operator representation with coinciding block structure of the Krein
space metric (involution operator). In the case of the Squire
equation we showed that the corresponding spectral problem is
connected with a ${\cal PT}-$symmetric eigenvalue problem by a
rescaling and Wick rotation\footnote{It is clear that, apart from
the Squire equation, there will exist other hydrodynamic equations
which can be structurally identified as Wick-rotated ${\cal
PT}-$symmetric systems in Krein spaces.}.

Based on recent results on the spectrum of the Squire equation, we
performed a qualitative analysis of the ${\cal PT}-$symmetric
quantum mechanical interpolation model for arbitrary square well
widths (cut-offs) $2b$. This allowed us to trace the emergence of
the Herbst limit with its empty spectrum as a spectral singularity
and to fit our results to those of the Bender-Boettcher equation
over the real line. We obtained a rich structure of multiple
spectral phase transitions from purely real eigenvalues to pairs of
complex conjugate ones --- as it was to expect for spectral problems
in Krein spaces.

A deeper insight into the Herbst-box spectrum and a possible
extension of the present results to ${\cal PT}-$sym\-met\-ric
Hamiltonians of the type $H_{M,N}=-\partial_x^2+x^{2M}(ix)^N, \
M,N=1,2,3,\ldots$ over square wells can probably be achieved by
representing the characteristic determinant $\Delta(E)$ in \rf{h3}
via Hadamar product representation of the Airy functions \cite{sal}
as a spectral determinant of Bethe-ansatz type
\cite{DDT-1,shin-cmp-1,DDT-n} and studying it by similar cocycle
functional equations as in Ref. \cite{voros}.

A question which was not touched in the present paper concerns the
orthogonality of the Herbst-box eigenfunctions. For the Squire
equation it is known that its eigenfunctions show a strong
non-orthogonality \cite{Step-1,RSH} (due to the non-normality of the
Squire operator) for eigenvalues in the vicinity of the branch point
center of the "Y" (pseudo-spectral techniques
\cite{Red-1,RSH,pseudo-spec} play an important role in this case).
Our above considerations indicate on a link of this issue with the
forming spectral singularity $|\partial_\nu E(\nu\approx -1,b\gg
1)|\gg 1$ in the vicinity of the almost vertical segments of the
purely real branch in the spectrum of the ${\cal PT}-$symmetric
interpolation model.

Finally, we would like to note two issues which seem of relevance
for future considerations. The first one is in developing efficient
mathematical tools to find the hyper-surfaces in parameter space
where spectral phase transitions of the real-to-complex type
occur\footnote{A two-step method similar in spirit was successfully
used, e.g., in higher-dimensional gravitational models to obtain the
stability regions in the moduli (parameter) space of these models
(step one: find  the critical hyper-surfaces; step two: identify the
stability/instability properties of the model aside of these
hyper-surfaces) \cite{grav}.}. Knowing these hyper-surfaces, one
would know the boundaries which separate the parameter space regions
with unbroken ${\cal PT}-$symmetry from regions with spontaneously
broken ${\cal PT}-$symmetry. In case of $\alpha^2-$dynamos the
corresponding knowledge would allow for a more precise prediction of
configurations with tendency to magnetic field reversals. The second
issue concerns methods for solving inverse spectral problems in
Krein spaces. Such methods would be extremely helpful for the data
analysis of the dynamo experiments which are planned for the near
future at seven sites around the world \cite{exp}.

\section*{Acknowledgements}

We thank G. Gerbeth, H. Langer, K.-H. R\"adler and C. Tretter for
useful comments. The project was supported by the German Research
Foundation DFG, grant GE 682/12-2, (U.G., F.S.) and by GA AS \v{C}R,
grant Nr.~1048302, (M.Z.).

\appendix
\section{A few comments on the physics of MHD $\alpha^2-$dynamos\label{dynamo-physics}}

The dynamo operator $\hat H_l[\alpha]$ originates from the MHD
mean-field induction equation (cf. \cite{krause1})
\begin{equation}\label{a1}
\partial_t \vect{B}=\bsym{\nabla \times }(\alpha \vect{B})
+\nu_m \Delta \vect{B}
\end{equation}
for the magnetic field $\vect{B}$. This equation results from
averaging over small scale turbulences in the velocity field of the
electrically conducting fluid (or plasma) which drives the dynamo.
The helical turbulence function $\alpha(x)$ (also called
$\alpha-$profile\footnote{In general setups, $\alpha$ is not a
scalar function but a tensor \cite{krause1}.}) encodes the net
effect of the small scale physics on the large scale (mean) magnetic
field $\vect{B}$. For certain topologically non-trivial helical
velocity and $\vect{B}-$field configurations an inverse cascade
effect occurs which induces an energy transfer from small-scale
structures to large-scale structures (inverse to the energy transfer
in usual turbulence cascades where the energy is pumped from
large-scale structures into smaller structures until it finally
dissipates and transforms into thermal energy). For sufficiently
strong inverse cascade effects the advection term $\bsym{\nabla
\times }(\alpha \vect{B})$ starts to dominate over the diffusion
term $\nu_m\Delta \vect{B}$ ($\nu_m$ is the magnetic diffusivity)
and the magnetic field strength starts to grow exponentially. This
kinematic dynamo effect (growing $ \vect{B}-$field for a given
velocity field of the fluid) is followed by a saturated dynamo
regime where a balance between the dynamo effect and the
back-reaction of the induced magnetic field on the velocity field
(via Navier-Stokes equation) prevents a further growth of the field
strength $ \vect{B}$. For completeness, we note that an MHD dynamo
is an open system in which part of the kinetic energy of the
conducting fluid (or plasma) transforms into magnetic field energy.

The dynamo eigenvalue problem
\be{a2} \hat H_l[\alpha]\phi_{l,n}=\lambda_{l,n}\phi_{l,n},\qquad \phi_{l,n}(t) \sim \exp
\lambda_{l,n}\,t
\ee
follows from the induction equation \rf{a1} via a double
decomposition: decomposing the $\vect{B}-$field into poloidal and
toroidal components (what leads to the two-component vector
structure of $\phi_{l,n}$) and expanding them further into spherical
harmonics. In the simplest (toy model) case of a spherically
symmetric dynamo configuration the corresponding modes decouple
completely and one arrives at the spherical $l-$mode projection
\rf{d1}, \rf{a2} (the subscript $n$ denotes the radial mode number).

Up to now only a single exactly solvable $\alpha^2-$dynamo model is
known --- the model with constant $\alpha-$profile \cite{krause1}.
Its spectrum is discrete, real \cite{KHR-2}, bounded above and,
depending on the value of $\alpha$, it is either completely negative
(for $\alpha$ below a critical $\alpha_c:\ \alpha<\alpha_c$) or it
contains a finite number of positive eigenvalues $\lambda_{l,n}>0$.
The dynamo effect is dominated by these latter eigenmodes. In
practice, it usually suffices to concentrate the analysis on the
dominating upper most growing mode (or a few of the upper most
modes) for dipole (l=1) and quadrupole (l=2) configurations. (There
exist no ``$s-$wave" $\alpha^2-$dynamos with $l=0$
\cite{krause1,GS-jmp1}.)

The spectral properties of the dynamo operator $\hat H_l[\alpha]$
are becoming much richer for inhomogeneous $\alpha-$profiles
$\alpha(r)\neq \const$, when real-to-complex transitions occur ---
as discussed in Subsection \ref{crossings} and shown in Fig.
\ref{fig-dynamo}.

\end{document}